\documentclass{article}

\usepackage{amsmath,amssymb,amsfonts,dcolumn,color,graphicx,graphics,latexsym,placeins,epsfig}
\usepackage{graphicx}  
\usepackage{amsmath}   
\usepackage{amssymb}   
\usepackage{bm} 
\usepackage{dcolumn}
\usepackage{color}
\usepackage{footmisc}
\usepackage{mathrsfs}
\usepackage{amsfonts}
\usepackage{varioref}
\usepackage{slashed}
\usepackage{footnote}
\usepackage[belowskip=-17pt,aboveskip=15pt]{caption}
\usepackage{amsmath}
\usepackage{makeidx}
\usepackage{amssymb}
\usepackage{graphicx}
\usepackage{bmpsize}
\usepackage{multicol}
\usepackage{url}
\usepackage{ragged2e}

\usepackage[margin=1.5in]{geometry}
\usepackage{graphicx}
\usepackage{dcolumn}
\usepackage{amsmath,amssymb,mathtools}
\usepackage{epstopdf}
\usepackage{verbatim}
\usepackage[colorlinks=true, citecolor=blue, urlcolor = blue, linkcolor= blue, 
bookmarks=true]{hyperref}
\usepackage[utf8]{inputenc}
\usepackage{amsmath}
\def\e{\epsilon}
\def\p{\partial}

\def\be{\begin{eqnarray}}
\def\ee{\end{eqnarray}}

\def\e{\epsilon}
\def\p{\partial}
\labelformat{section}{Section #1} 
\labelformat{subsection}{Section #1} 
\labelformat{subsubsection}{Section #1}
\labelformat{subsubsubsection}{Section #1}
\labelformat{eqnarray}{Eq.~(#1)} 
\labelformat{figure}{Fig.~#1} 
\labelformat{subfigure}{Fig.~\thefigure#1} 
\labelformat{table}{Tab.~#1} 
\labelformat{appendix}{Appendix #1}
\vspace{25pt}
\title{\bf Effect of $\xi R\phi^2$ non-minimal coupling on gravitational light bending}
\author{\bf Ayan Kumar Naskar,\footnote{meayannaskar@gmail.com } \,\, Avijit Sen Majumder, \footnote{senmajumderavijit@gmail.com} \,\, and \,\,  Sourav Bhattacharya\footnote{sbhatta.physics@jadavpuruniversity.in}\\
\small{Relativity and Cosmology Research Centre, Department of Physics, Jadavpur University, Kolkata 700 032, India}\\}

\begin{document}
\maketitle
\begin{abstract}
\noindent  We investigate the bending of massless fields by a massive object in the presence of a curvature-scalar $\sqrt{-g}\xi R \phi^2$ non-minimal coupling  up to one loop, using the perturbative quantum gravity computations.  It is well known that without such coupling a self interacting scalar field theory cannot be renormalised in the presence of gravity. The massive object is modelled by a massive scalar  $\phi$, and it is assumed to be non-relativistic, e.g., a star. We compute the 2-2 scattering of massless scalar and photons off this object via graviton exchanges. Assuming both   $\xi$ and the bending angle to be small, we use the eikonal approximation to compute the angle up to ${\cal O}(\xi G^2)$. At  tree level $({\cal O}(\xi G))$ we find  no bending, and hence the ${\cal O}(\xi G^2)$ result happens to be leading in this case. The non-minimal vertices are qualitatively different from that of the standard minimal ones, e.g. $ \sqrt{G} h_{\mu\nu} T^{\mu\nu}$, as  the former contains explicit momenta of the gravitons instead of the scalar, complementing the second. The bending angle is found to behave like $\sim b^{-7}$, where $b$ is the impact parameter. We have emphasised the qualitative differences of our results  from that of the well studied minimal  case.
\end{abstract}

\section{Introduction}\label{S1}

It is well accepted since long that the theory of gravity, when treated as a perturbative quantum field theory, is not renormalisable~\cite{Utiyama, Weinberg:1965nx, Pagels:1966zza, vanDam:1970vg, Zakharov:1970cc, Iwasaki:1971vb, Boulware:1972yco, tHooft:1974toh, deser1, deser2, deser3, Stelle, Voronov, Goroff:1985th,  Odintsov4, Odintsov5, PRD:87, Lavrov}, and references therein. Despite numerous efforts  (see e.g.~\cite{Mukhanov, Keifer, Shapiro} for vast reviews), there exists no complete or satisfactory quantum theory of gravitation to date. 

{\color{black}Even though non-renormalisable, it is generally believed   that for energy scales much below the Planck mass, the first few orders of perturbative quantum gravity computations might lead to interesting physical predictions, e.g.~\cite{Hiida:1972xs, Barker:1975ae, Donoghue:1993eb, Donoghue:1994dn, Muzinich:1995uj,  Hamber:1995cq, MODANESE1995697, Akhundov:1996jd, Donoghue_1996, Bjerrum-Bohr:2002aqa, Bjerrum-Bohr:2002fji, Bjerrum-Bohr:2002gqz, Bern:2002kj, Akhundov:2006gh, Holstein:2008sx,Toms:2010vy, Bjerrum-Bohr:2015vda, Malta2015ComparativeAO, Frob:2016xte,Ulhoa_2017, Olyaei:2018asy, deBrito:2020wmp, Frob:2021mpb, Donoghue:2017pgk, Majumder:2025sht, Majumder:2026hkb, Toms:2008dq, Toms:2009zz, Toms:2009vd, Toms:2011zza, Majumder:2025gou} and references therein.  At the Planck scale, on the other hand (such as near the big bang singularity), one expects a non-perturbative or a complete quantum theory of  gravity to be appropriate. The recent discovery of gravitational waves has also led to  proposals for detecting some resummed quantum gravity effects in model dependent or independent manner through certain observable quantities, e.g.~\cite{Calcagni:2019ngc, Garcia-Chung:2022pdy} and references therein.   As of  the perturbative quantum gravity  predictions we mentioned above, we may hope that they  will be  testable, hopefully in a not too distant future,  with the modest  expectation that they will one day be successfully embedded in  a complete theory of quantum gravity. One of the most studied, and hopefully testable in future,  aspects in this framework is the long range two-body gravitational potential and the bending of light.} See also  e.g.~\cite{Donoghue:1995cz, Burgess:2003jk, Goldberger:2004jt, Foffa:2016rgu, Levi:2018nxp, Cheung:2018wkq, Bern:2020buy, Ivanov:2022qqt, Almeida:2024uph, Almeida:2024cqz, Trestini:2024mfs} and references therein for an effective field theory description of gravity. We further refer our reader to e.g.~\cite{Seery:2007we, Woodard:2014jba, Oda:2015sma, Moss:2014nya, Shapiro:2015ova, Saltas:2015vsc, Buchbinder, Arbuzov:2021yai} and references therein for discussion on perturbative quantum gravity in curved spacetimes.

In this work we wish to confront the curvature-scalar  $\sqrt{-g}\xi R \phi^2/2$ non-minimal coupling from the point of view of perturbative quantum gravity.  It is well known that such coupling becomes {\it necessary} while renormalising a  self interacting quantum  scalar field theory in curved spacetimes~\cite{Parker:2009uva}. For  a conformal scalar ($\xi=1/6$ and $m^2=0$), classical hairy back hole solutions can be seen in~\cite{Bocharova:1970skc, Bekenstein:1974sf, Martinez:2002ru, Bhattacharya:2013hvm, Bravo-Gaete:2025vyd}. See also~\cite{Vagnozzi:2025yfm} for discussion on Hawking radiation in the context of non-minimal coupling. Note that the $\sqrt{-g}\xi R \phi^2/2$ coupling generates  two scalar-$n$ graviton vertices, containing the momenta carried by the graviton lines explicitly, but  not of the scalar lines.  This  is qualitatively opposite to the  usual matter-graviton vertices (like $\kappa h_{\mu\nu} T^{\mu\nu}$), where  the momenta carried by the matter field, and {\it not} of the graviton, appear explicitly. Computation of the one loop two body long range gravitational potential for such coupling was reported  recently in~\cite{Majumder:2026hkb, Majumder:2025gou}.

Specifically, we wish to investigate in this work the effect of $\xi$ on the bending of massless scalar and photons at the leading $({\cal O}(\xi G))$ and next to the leading $({\cal O}(\xi G^2))$ order. We refer our reader to~\cite{Bjerrum-Bohr:2014zsa, Bai:2016ivl, Bjerrum-Bohr:2017dxw, PDV, Bastianelli:2021nbs} perturbative quantum gravity computations for light bending for the minimal case. It turns out that the ${\cal O}(\xi G^2)$ result is the leading one in this case. We will also see towards the end of this paper that the aforementioned feature of the non-minimal vertices will bring in a counter-intuitive feature in the bending angle. We will assume $\xi$ to be small, but will leave it arbitrary otherwise. \\

\noindent
We will work with the mostly positive signature of the metric in $(3+1)$ spacetime dimensions. The incoming (outgoing) momenta in the 2-2 scattering process will always be denoted by $k_1, k_2$ ($k'_1, k'_2$) respectively. $k_1$ and $k'_1$ will be associated with the massive scalar with non-minimal coupling, whereas $k_2, k'_2$ will stand for the massless fields, for which {\it no} non-minimal coupling will be assumed. The external momenta are on shell, i.e., $k_1^2= {k'}_1^2= -M^2$ and $k_2^2= {k'}_2^2=0$. In addition, $k_1$ and $k'_1$ will be taken to be non-relativistic (e.g. a star).  The transfer momentum ($k_1-k'_1$, or $k'_2-k_2$) will always be denoted by $q$.  For symmetrisation, we will use the notation : $X_{(\alpha}Y_{\beta)}= X_{\alpha}Y_{\beta}+X_{\beta}Y_{\alpha}$.

\section{The basic ingredients}\label{S2}
Let us begin by briefly noting the basic framework we will be working in. The action of our theory reads
\begin{eqnarray}
    \begin{split}
         S =& \frac{2}{\kappa^2}\int d^4 x \sqrt{-g} R -\frac12  \int d^4 x \sqrt{-g}\left[ g^{\mu\nu}(\nabla_{\mu}\phi)(\nabla_{\nu}\phi) + \xi R \phi^2 + m^2\phi^2 \right]-\frac12  \int d^4 x \sqrt{-g}g^{\mu\nu}(\nabla_{\mu}\varphi)(\nabla_{\nu}\varphi) \\ 
         &  -  \frac14\int d^4 x \sqrt{-g}\  g^{\mu\alpha}g^{\nu\beta}F_{\mu\nu} F_{\alpha\beta}, 
\label{qg1}
    \end{split}
\end{eqnarray}
where $\kappa^2 = 32 \pi G$. We wish to  compute below scattering of the massless scalar and photons via graviton exchange off the massive scalar up to ${\cal O}(\xi G^2)$.

We will essentially work in the weak gravity regime where the metric can be decomposed into the Minkowski background plus a quantum fluctuation over it, so that
\begin{eqnarray}
&& g_{\mu\nu}=\eta_{\mu\nu}+\kappa h_{\mu\nu}, \quad  g^{\mu\nu}=\eta^{\mu\nu}-\kappa h^{\mu\nu}+\kappa^2 h^{\mu}{}_{\alpha} h^{\alpha \nu} +\cdots, \quad 
\sqrt{-g} = 1 +\frac{\kappa h}{2} + \frac{\kappa^2 h^2}{8} -\frac{\kappa^2 }{4} h_{\mu\nu} h^{\mu\nu} + \cdots \nonumber\\ &&
\Gamma^{\mu}_{\nu\rho}= \frac{\kappa}{2} \eta^{\mu\alpha}\left(\p_{\nu}h_{\rho \alpha} + \p_{\rho} h_{\nu\alpha}-\p_{\alpha} h_{\nu\rho} \right)-\frac{\kappa^2}{2} h^{\mu\alpha}\left(\p_{\nu}h_{\rho \alpha} + \p_{\rho} h_{\nu\alpha}-\p_{\alpha} h_{\nu\rho} \right) + \frac{\kappa^3}{2} h^{\mu\beta} h_{\beta}{}^{\alpha}\left(\p_{\nu}h_{\rho \alpha} + \p_{\rho} h_{\nu\alpha}-\p_{\alpha} h_{\nu\rho} \right) +\cdots \nonumber\\
\label{qg2}
\end{eqnarray}
We will take $k_1(k'_1)$ to be the external momenta of the massive scalar ($k_1^2=k'^2_1=-M^2$), whereas $k_2(k'_2)$ will be the same for the massless fields ($k_2^2=k'^2_2=0$). The transfer momentum will be denoted by $q=k_1-k'_1=k'_2-k_2$.

The graviton propagator in the de Donder gauge,
$$\p_{\mu}\left(h^{\mu}{}_{\nu}-\frac12 \delta^{\mu}_{\nu} h \right)=0$$
reads,
{\color{black}
\begin{eqnarray}
	D_{\mu\nu,\rho\sigma}(q)
	=
	\frac{i}{q^2+i\epsilon}
	\left[
	\frac{1}{2}
	\left(
	\eta_{\mu\rho}\eta_{\nu\sigma}
	+
	\eta_{\mu\sigma}\eta_{\nu\rho}
	\right)
	-
	\frac{1}{d-2}
	\eta_{\mu\nu}\eta_{\rho\sigma}
	\right],
	\label{qg4}
\end{eqnarray}
where $d=4-\varepsilon$ is taken for the sake of dimensional regularisation.}

\begin{figure}[h!]
\begin{center}
    \includegraphics[width=0.18\linewidth]{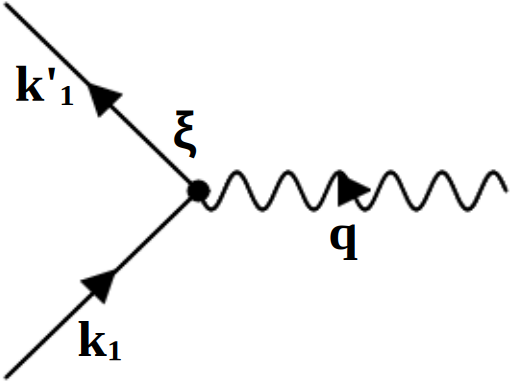} 
    \qquad \qquad \qquad \qquad \includegraphics[width=0.15\linewidth]{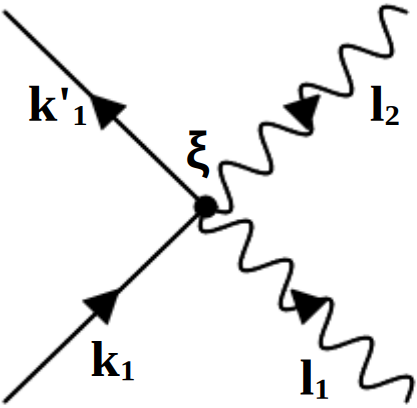} 
  \caption{\small \it The one and two graviton-two scalar non-minimal vertices. The dark circle on the junction asserts that these vertices are non-minimal, in contrast to the minimal ones.   }
  \label{fnm}
\end{center}
\end{figure}

\noindent
Let us now summarise the vertex functions we will be needing for our purpose. For the standard minimal ($\xi=0$) interactions, the one graviton-massive two scalar and two graviton-massive two scalar vertices respectively read~\cite{Holstein:2008sx}
\begin{eqnarray}
&& V_{\text{spin-0}}^{(1)\,\mu \nu} (k,k',m)= -\frac{i\kappa}{2} \left[k^{\mu}k'^{\nu}+k'^{\mu}k^{\nu}-\eta^{\mu\nu}(k\cdot k' + m^2)  \right], \nonumber \\
 &&       V_{\text{\:spin-0}}^{(2)\,\mu \lambda \rho \sigma} (k,k',m) = i\kappa^2 \Big[\Big\{ I^{\mu \lambda \alpha \nu}  I^{\rho\sigma \beta}{}_{\nu} -\frac14 \big(\eta^{\mu \lambda}I^{\rho\sigma \alpha \beta} +\eta^{\rho \sigma}I^{\mu \lambda \alpha \beta}  \big)  \Big\} k_{(\alpha}k'_{\beta)} -\frac12 \Big(I^{\mu \lambda \rho \sigma} -\frac12 \eta^{\mu \lambda}\eta^{\rho\sigma} \Big) (k\cdot k' + m^2)   \Big] \nonumber\\
     \label{qg21}
\end{eqnarray}
where, 
$$I_{\mu\nu\lambda \rho}= \dfrac12 \Big( \eta_{\mu\lambda} \eta_{\nu\rho} + \eta_{\mu\rho} \eta_{\nu\lambda}\Big)$$
The momenta appearing in Eqs.~\ref{qg21} are carried by the scalar.
\noindent
The one graviton-two scalar and two graviton-two scalar {\it non-minimal} vertices respectively read, \ref{fnm}~\cite{Majumder:2026hkb},
\begin{eqnarray}
&&    V^{\xi}_{\mu \nu} (\xi \kappa, k) = -i\xi \kappa k^2 \eta_{\mu\nu},\nonumber\\
&&    V_{\xi}^{\mu \nu; \rho \sigma} (l_1,l_2) = - \frac{i \xi \kappa^2 }{4}  \Big[ (l_1^2 +l_2^2) (\eta^{\mu \nu} \eta^{\rho \sigma} - 4 \eta^{\mu \rho} \eta^{\nu \sigma}) - 2 ( l_1^{\sigma} l_2^{\nu } + l_1^{\nu} l_2^{\sigma } ) \eta^{\mu \rho} + 6 l_1.l_2   \eta^{\mu \rho } \eta^{ \nu \sigma }   \Big]
    \label{qg21a}
\end{eqnarray}
The momenta appearing in the above equations are carried by gravitons. We also note the three graviton vertex,
\begin{eqnarray}
    \begin{split}
    \label{qg25}
        V^{(3)\mu\nu}_{\alpha\beta \gamma\delta}(k, q) =& -\frac{i\kappa}{2} \Big[  P_{\alpha\beta \gamma\delta}\Big(k^{\mu}k^{\nu}+(k-q)^{\mu}(k-q)^{\nu}+q^{\mu}q^{\nu} -\frac32 \eta^{\mu\nu}q^2\Big) +2 q_{\lambda} q_{\sigma} \Big(I_{\alpha\beta}{}^{\sigma\lambda}I_{\gamma\delta}{}^{\mu\nu}+I_{\gamma\delta}{}^{\sigma\lambda}I_{\alpha\beta}{}^{\mu\nu} \\
        & -I_{\alpha\beta}{}^{\mu\sigma}I_{\gamma\delta}{}^{\nu\lambda}- I_{\gamma\delta}{}^{\mu\sigma}I_{\alpha\beta}{}^{\nu\lambda}  \Big) +\Big\{q_{\lambda} q^{\mu}\Big(\eta_{\alpha\beta} I_{\gamma\delta}{}^{\nu\lambda}+\eta_{\gamma\delta} I_{\alpha\beta}{}^{\nu\lambda} \Big)   + q_{\lambda} q^{\nu}\Big(\eta_{\alpha\beta} I_{\gamma\delta}{}^{\mu\lambda}+\eta_{\gamma\delta} I_{\alpha\beta}{}^{\mu\lambda} \Big) \\
        & - q^2 \Big(\eta_{\alpha\beta} I_{\gamma\delta}{}^{\mu\nu}+\eta_{\gamma\delta} I_{\alpha\beta}{}^{\mu\nu}\Big) - \eta^{\mu\nu}q_{\sigma}q_{\lambda}\Big(\eta_{\alpha \beta}I_{\gamma\delta}{}^{\sigma\lambda}+ \eta_{\gamma\delta}I_{\alpha \beta}{}^{\sigma\lambda}\Big)  \Big\} + \Big\{-2q_{\lambda} \Big(I_{\alpha\beta}{}^{\lambda \sigma} I_{\gamma\delta \sigma}{}^{\nu} (k-q)^{\mu} \\
        & +I_{\alpha\beta}{}^{\lambda \sigma} I_{\gamma\delta \sigma}{}^{\mu} (k-q)^{\nu} +  I_{\gamma\delta}{}^{\lambda \sigma}I_{\alpha\beta\sigma}{}^{\nu}k^{\mu}+ I_{\gamma\delta}{}^{\lambda \sigma}I_{\alpha\beta\sigma}{}^{\mu}k^{\nu} \Big) +q^2 \Big(I_{\alpha\beta\sigma}{}^{\mu}I_{\gamma\delta}{}^{\nu\sigma}+ I_{\gamma\delta\sigma}{}^{\mu}I_{\alpha\beta}{}^{\nu\sigma} \Big) \\
        & +\eta^{\mu\nu}q_{\sigma}q_{\lambda}\Big(I_{\alpha\beta}{}^{\lambda\rho}I_{\gamma\delta \rho}{}^{\sigma}+ I_{\gamma\delta}{}^{\lambda\rho}I_{\alpha\beta\rho}{}^{\sigma} \Big)  \Big\} + \Big\{\Big(k^2+(k-q)^2 \Big) \Big(I_{\alpha\beta}{}^{\mu\sigma}I_{\gamma\delta\sigma}{}^{\nu} + I_{\gamma\delta}{}^{\mu\sigma}I_{\alpha\beta\sigma}{}^{\nu} 
         -\frac12 \eta^{\mu\nu}{\cal P}_{\alpha\beta\gamma\delta}\Big)\\ & - \Big(I_{\gamma\delta}{}^{\mu\nu}\eta_{\alpha\beta}k^2+I_{\alpha\beta}{}^{\mu\nu}\eta_{\gamma\delta}(k-q)^2  \Big)  \Big\} \Big]
    \end{split}
\end{eqnarray}

\noindent
The two photon-one graviton and  the two photon-two graviton vertices respectively read~\cite{Holstein:2008sx},
\begin{eqnarray}
     &&    V^{\text{spin-1}\,(1)}_{\alpha,\beta; \mu\nu}(k_1,k_2) = - \dfrac{i \kappa}{  2}   \eta_{\mu \nu} \left(k_1 \cdot k_2 \eta_{\alpha \beta} - k_{1 \beta} k_{2 \alpha} \right) + i \kappa   I_{\mu \nu \kappa \lambda} \left[k_1 \cdot k_2 {I_{\alpha \beta}}^{\kappa \lambda}   + \frac{1}{2} \big(k_1^\kappa k_2^\lambda + k_1^\lambda k_2^\kappa\big) \eta_{\alpha \beta} - \big(k_1^\kappa k_{2 \alpha} \delta_\beta^\lambda + k_2^\kappa k_{1 \beta} \delta_\alpha^\lambda \big)\right], \nonumber\\ &&
         V^{\text{spin-1}\,(2)}_{\alpha,\beta; \mu\nu, \rho \sigma}(k_1,k_2) =  {i \kappa^2 \over 2}  P_{\mu \nu \rho \sigma} \left(k_1 \cdot k_2 \eta_{\alpha \beta} - k_{1 \beta} k_{2 \alpha} \right)  -  i \kappa^2 \Big\{  {I_{\mu \nu}}^{\kappa \delta} {I_{\rho \sigma \delta}}^{\lambda} +  {I_{\rho \sigma}}^{\kappa \delta} {I_{\mu \nu \delta}}^{\lambda} \nonumber\\ && -  \frac{1}{2} \Big(\eta_{\mu \nu} {I_{\rho \sigma}}^{\kappa \lambda}  +  \eta_{\rho \sigma} {I_{\mu \nu}}^{\kappa \lambda}\Big)  \Big\}  \Big[k_1 \cdot k_2 I_{\alpha \beta\kappa \lambda} + \frac{1}{2} \big(k_{1\kappa} k_{2\lambda} + k_{1\lambda} k_{2\kappa}\big) \eta_{\alpha \beta} 
         - \big(k_{1\kappa} k_{2 \alpha} \delta_{\beta \lambda} + k_{2\kappa} k_{1 \beta} \delta_{\alpha \lambda} \big)\Big]\nonumber \\ &&  - \frac{i \kappa^2 }{2}  \Big({I_{\mu \nu}}^{\eta \theta} {I_{\rho \sigma}}^{\kappa \lambda} + {I_{\rho \sigma}}^{\eta \theta} {I_{\mu \nu}}^{\kappa \lambda} \Big)   \left[k_{1 \eta} \eta_{\alpha \kappa} (k_{2 \theta} \eta_{\beta \lambda} -  k_{2 \lambda} \eta_{\beta \theta} ) +  k_{2 \eta} \eta_{\beta \kappa} (k_{1 \theta} \eta_{\alpha \lambda}  -  k_{1 \lambda} \eta_{\alpha \theta} )  \right],
         \label{l0}
\end{eqnarray}
We work  in the Lorentz gauge, $\p_{\mu}A^{\mu}=0$.\\

\noindent
{\color{black} We will evaluate bending angles for  massless fields via the eikonal phase, which we wish to review very briefly now, following~\cite{Bjerrum-Bohr:2017dxw, PDV, Bastianelli:2021nbs} and references therein.
We assume that the bending angle is small, and also  the momentum transfer  $(q)$ between the massive and massless fields is small compared to their individual momenta.  Hence the scattering matrix introduces a phase only. We choose a frame  where the massive object is at rest initially. We assume that it  is   heavy (such as a star or a black hole), and  hence by the virtue of small momentum transfer, we may ignore its kinetic energy compared to its rest mass energy after the scattering as well. Thus, if the initial and final four momenta of this object are respectively $k_1$ and $k'_1$, for  $q=k_1-k'_1 $, we have $q^0 \ll |\vec{q}|$.
In addition, one can show in such an effectively static heavy mass limit,
$$q^2=\vec{q}_{\perp}^{2}\left[ 1+\mathcal{O}\left(\dfrac{\vec{q}_{\perp}^2}{E^2}, \ \dfrac{\vec{q}_{\perp}^2}{M^2}\right)\right],$$
where $\vec{q}_{\perp}$ is the transfer momentum orthogonal to the direction of propagation. Since by our assumption $|\vec{q}| \ll E, M$, we will take below  $q^2\simeq \vec{q}_{\perp}^{2}$ in the   scattering  amplitudes, known as the {\it standard eikonal replacement}. Thus, after evaluation of  a scattering amplitude in the usual  Lorentz invariant manner, in the present scenario one  evaluates it on the ($d-2$)-dimensional impact parameter space orthogonal to the direction of propagation.  We have the Fourier transform 
$$\widetilde{\mathcal{M}}(b)= \frac{1}{4EM}\int \frac{d^{d-2}\,\vec{q}_{\perp}}{(2\pi)^{\color{black}{d-2}}} \mathcal{M}(\vec{q}_{\perp})\;e^{i\vec{q}_{\perp}\cdot\vec{b}}.$$ 
The scattering matrix can be identified  as  $S \sim e^{i\chi}$, where $\chi$ is the eikonal phase. This yields $ i \widetilde{\mathcal{M}}(b)= e^{i\chi(b)} -1$. The bending angle can then be obtained from the formula,
\begin{equation}
	\theta_b =- \frac{1}{E}\frac{\partial\chi(b)}{\partial b}. \label{eqeikonal}
\end{equation}

\noindent
Finally, we also note that any non-vanishing value of the long range bending angle must correspond to $ \mathcal{M}(q)$'s non-analytic dependence on the transfer momentum $q$, which will be our main focus in the following computations. }With these equipments, we are now ready to go into the calculation of scattering.
\section{The scattering amplitudes and the bending angles}\label{S3}
\begin{figure}[h]
         \hspace{1cm}\begin{minipage}[b]{0.3\linewidth}
             \centering

             \includegraphics[width=0.7\linewidth]{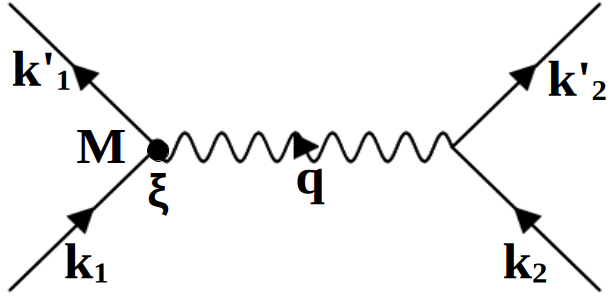}
          \end{minipage} 
         \begin{minipage}[b]{0.3\linewidth}
             \centering
             \includegraphics[width=0.5\linewidth]{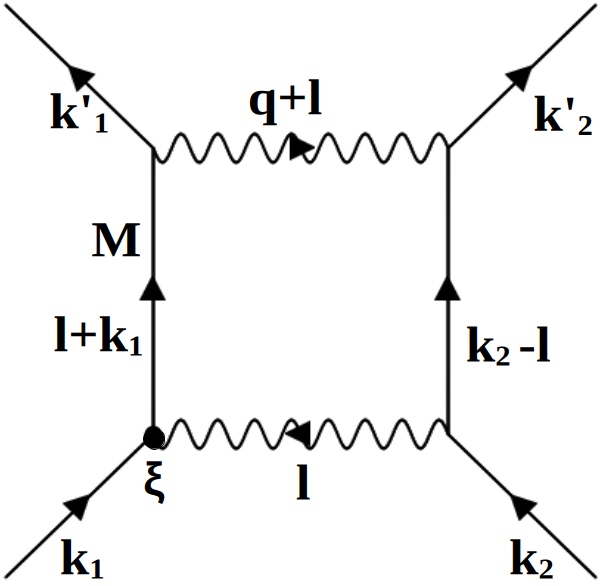}
         \end{minipage}
         \begin{minipage}[b]{0.3\linewidth}
             \centering
             \includegraphics[width=0.5\linewidth]{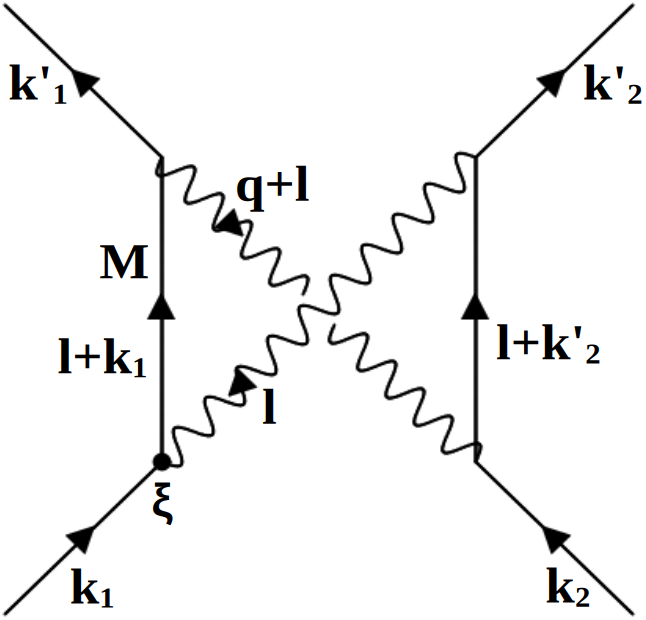}
          \end{minipage} 
           \caption{\it \small The tree, ladder, and cross-ladder diagrams at linear order in the non-minimal coupling parameter $\xi$.  The ladder and cross-ladder diagrams have two sub-categories each, depending on the placement of the $\xi$-vertex, denoted by the thick circle.}
           \label{tr-l-crl}
    \end{figure}

{\color{black} Let us now evaluate the leading and next to the leading order  scattering processes depicted in \ref{tr-l-crl}, \ref{triangle2grav}. We will take the on shell relations, $k_1^2= {k'}_1^2=-M^2$, $k_2^2={k'}_2^2=0$ for the external fields. We will also use below, $k_1\cdot q= q^2/2=-k'_1\cdot q$, and $k_2\cdot q= -q^2/2=-k'_2\cdot q$. }
\subsection{The massive scalar-massless scalar scattering}

\subsection*{\small a) The tree diagram : }

\noindent
We begin by considering the tree diagram, the first of \ref{tr-l-crl}. The amplitude reads 
\begin{eqnarray}
&& i  \mathcal{M}^{\text{spin-0-spin-0}}_{\text{tree}}=-\frac{i}{2} (-i\kappa \xi) \left( - \frac{i\kappa}{2}\right) \frac{q^2 \eta^{\alpha\beta}{\cal P}_{\mu\nu\alpha\beta}\left(k_2^{\mu}{k'}^{\nu}_2+k_2^{\nu}{k'}^{\mu}_2-\eta^{\mu\nu}k_2\cdot k'_2\right) }{q^2}= -\frac{i\kappa^2\xi q^2}{2}\cdot 
\label{t1}
\end{eqnarray}
The Fourier transform of this amplitude is proportional to $\vec{\p}^2\delta^3(\vec{r})$, i.e. purely local,  and hence it does not contribute to any long range gravitational potential or to gravitational light bending. 

\subsection*{\small b) The ladder and the cross-ladder diagrams : }

\noindent
The ladder diagram is given by the second of \ref{tr-l-crl}. There are two sub-categories here depending upon whether the $\xi$-vertex 
is placed on the $k'_1$ or $k_1$ lines. For the first, we have 
\begin{eqnarray}
&& i  \mathcal{M}^{\text{spin-0-spin-0}}_{\text{ladder,1}}=\frac{\xi \kappa^4}{2^5}\eta_{\mu\nu}{\cal P}^{\mu\nu\alpha\beta}\left\{ k'_{2(\alpha} (k_2-l)_{\beta)} + \eta_{\alpha\beta} k'_2\cdot (l-k_2) \right\}   \nonumber\\&& \times
\frac{\left\{k'_{1(\lambda} (l+k_1)_{\rho)} - \eta_{\rho\lambda} k_1\cdot l  \right\} {\cal P}^{\lambda \rho \gamma \delta} \left\{  k_{2(\gamma} (k_2-l)_{\delta)} + \eta_{\gamma\delta} k_2\cdot l\right\} }
{l^2 (l-k_2)^2 [(l+k_1)^2+M^2]}  \cdot 
\label{t2}
\end{eqnarray}
The above amplitude has no terms non-analytic in the transfer momentum $q^2$, and hence has no contribution to light bending. The same conclusion holds for the second ladder diagram where the $\xi$-vertex is placed on the $k_1$ line.\\

\noindent
Let us now come to the cross-ladder diagram, which also has two sub-categories as of the ladder diagram. For the $\xi$-vertex is placed upon the $k'_1$ line, we have 
\begin{eqnarray}
&& i  \mathcal{M}^{\text{spin-0-spin-0}}_{\text{cross-ladder,1}}=\frac{\xi \kappa^4}{2^5}\eta_{\mu\nu}{\cal P}^{\mu\nu\alpha\beta}\left\{ k_{2(\alpha} (k'_2+l)_{\beta)} - \eta_{\alpha\beta} k_2\cdot l \right\}   \nonumber\\&& \times
\frac{\left\{k_{1(\lambda} (l+k_1)_{\rho)} - \eta_{\rho\lambda} k_1\cdot l  \right\} {\cal P}^{\lambda \rho \gamma \delta} \left\{  k'_{2(\gamma} (k'_2+l)_{\delta)} - \eta_{\gamma\delta} k'_2\cdot (l+k'_2)\right\} }
{l^2 (l+k'_2)^2 [(l+k_1)^2+M^2]},
\label{t3}
\end{eqnarray}
which also does not contribute to the long range potential and hence to the light bending. Likewise, the second sub-category does not contribute to the same. 
\begin{figure}[h]
	\hspace{0.55cm}
        \begin{minipage}[b]{0.25\linewidth}
            \centering
            \includegraphics[width=0.7\linewidth]{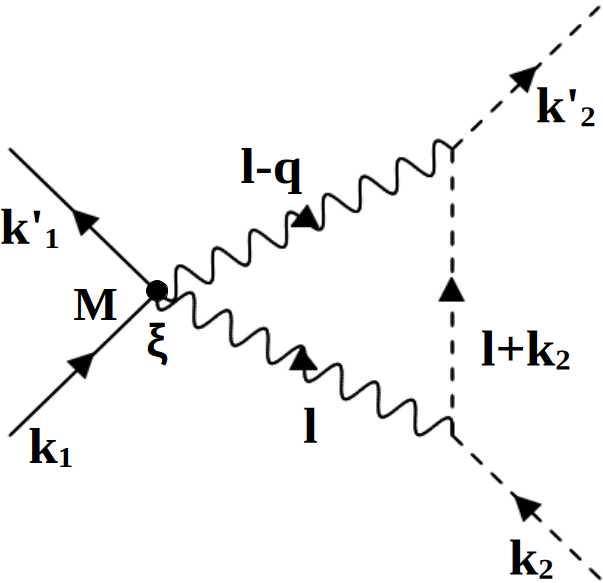}
        \end{minipage}
        \hspace{1.4cm}
        \begin{minipage}[b]{0.25\linewidth}
            \centering
            \includegraphics[width=0.7\linewidth]{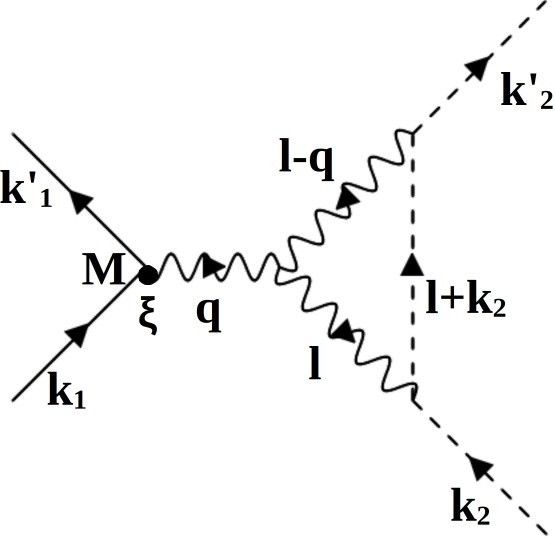}
         \end{minipage} 
        \hspace{1cm}
        \begin{minipage}[b]{0.25\linewidth}
            \centering
            \includegraphics[width=0.65\linewidth]{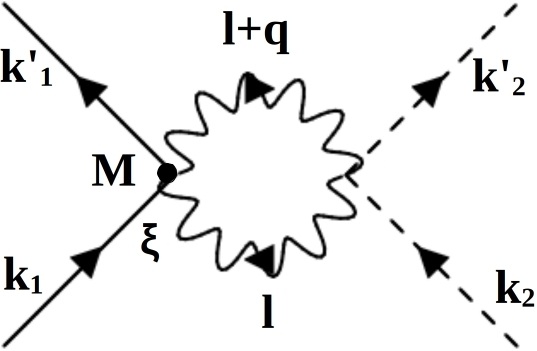}
        \end{minipage} \\
        \\ \\
        \begin{minipage}[b]{0.3\linewidth}
            \centering
            \includegraphics[width=0.68\linewidth]{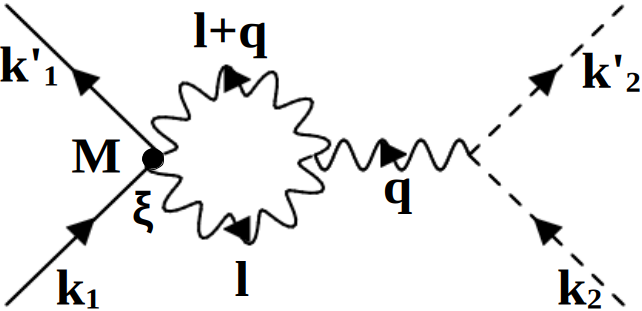}
        \end{minipage}
        \hspace{0.7cm}
        \begin{minipage}[b]{0.3\linewidth}
            \centering
            \includegraphics[width=0.7\linewidth]{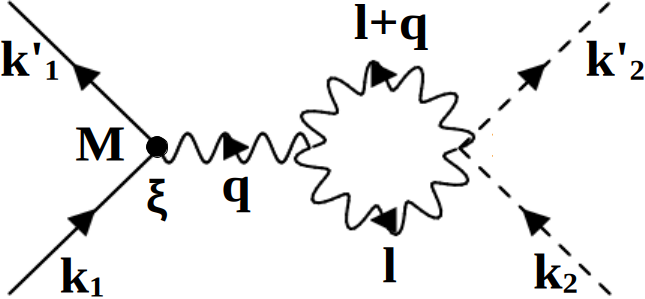}
         \end{minipage} 
        \hspace{0.7cm}
        \begin{minipage}[b]{0.3\linewidth}
            \centering
            \includegraphics[width=0.7\linewidth]{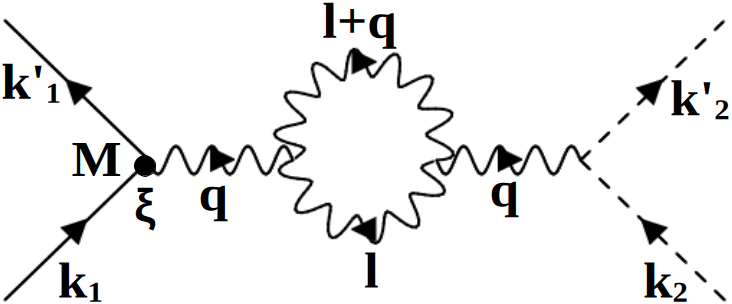}
        \end{minipage}\\
        \caption{\it \small The diagrams for massive-massless scalar, and massive scalar-photon scattering. The broken lines will consecutively represent the massless scalar field for this section, and the photon  in the next section. We have fixed momenta $(k_1,k_1')$ for the non-minimal  massive scalar represented by the solid lines. }
         \label{triangle2grav}
\end{figure}
\subsection*{\small c) The triangle diagrams :}
 The  Feynman amplitude for the first of \ref{triangle2grav} reads,
\begin{eqnarray}
	&&  i \mathcal{M}^{\text{spin-0-spin-0}}_{\text{triangle}} = \int \frac{d^{\color{black}{d}} l}{(2 \pi)^{\color{black}{d}}}  V^{\text{spin-0(2)}}_{\phi \lambda \rho \sigma (\xi)} (l,l+q) \frac{{- i \cal P}^{ \phi \lambda \mu \nu}}{l^2} \frac{{- i \cal P}^{ \rho \sigma \alpha \beta }}{(l+q)^2} V^{\text{spin-0(1)}}_{\mu \nu} (l+ k_2, k_2') V^{\text{spin-0(1)}}_{\alpha \beta} (k_2,l+k_2) \frac{-i}{(l+k_2)^2 }. \nonumber\\
	\label{l1'}
\end{eqnarray}
\noindent
{\color{black}{We will evaluate the loop integrals using dimensional regularisation in $d=4-\varepsilon$,  and will retain only the non-analytic dependence on the transfer momentum   $q^{\mu}$, relevant to long-range bending. Terms which are analytic in $q$  produce contact interactions and hence are not much  interesting physically. We will not demonstrate  renormalisation   here explicitly. Using the ingredients described in \ref{S2}, Eq.~\ref{l1'} can be evaluated by standard Feynman  integral technique to give 
\begin{eqnarray}
	&&  i \mathcal{M}^{\text{spin-0-spin-0}}_{\text{triangle}}= \frac{193 G^2 \xi  i }{6}   q^4 \ln q^2.
	\label{potamp2a}
\end{eqnarray}
Note  that the above amplitude is Lorentz invariant. We next use the eikonal approximation described towards the end of \ref{S2} to  obtain the corresponding bending angle,
\begin{eqnarray}
	\theta^{\text{spin-0-spin-0}}_{\text{triangle}} =& -\dfrac{1544 G^2 \xi }{E^2 \pi  b^7 M}\cdot
	\label{potamp21}
\end{eqnarray}
Note also that if we retain the $\mathcal{O}\left(\vec{q}_{\perp}^2/E^2, \ \vec{q}_{\perp}^2/M^2\right)$ corrections to the leading formula, $q^2\simeq  q_{\perp}^2$ (cf. discussion below Eq.~\ref{l0}), we would have produced corrections to the above  bending angle as
 $$\theta^{\text{spin-0-spin-0}}_{\text{triangle}} \to \theta^{\text{spin-0-spin-0}}_{\text{triangle}}\left(1+\mathcal{O}\left(1/(bE)^2, \ 1/(bM)^2\right)\right)$$
 We have thus ignored them as they are much subleading compared to the expression in Eq.~\ref{potamp21}. }\\

\noindent
 We also notice the peculiar appearance of the mass in the denominator. This corresponds to the absence of any explicit scalar momentum in the non-minimal vertex function, Eq.~\ref{qg21a}. This feature seems to be counterintuitive, and we will come to its possible physical implication towards the end of this letter. }
\subsection*{\small  d) The seagull diagram :}
The seagull diagram, given by the second  of \ref{triangle2grav}, yields the Feynman amplitude
\begin{eqnarray}
    \begin{split}
        i \mathcal{M}^{\text{spin-0-spin-0}}_{\text{seagull}} =& \int \frac{d^{\color{black}{d}} l}{(2 \pi)^{\color{black}{d}}} V_{\text{spin-0}(1)}^{\lambda \phi (\xi)} (q) \frac{-i \mathcal{P}_{\lambda \phi \mu \nu}}{q^2} V^{\mu \nu(3)} _{\alpha \beta \gamma \delta} (l-q,-q) \frac{-i \mathcal{P}^{\alpha \beta \psi \theta}}{l^2}  \frac{-i \mathcal{P}^{\gamma \delta \rho \sigma}}{(l-q)^2}  V^{\text{spin-0(1)}}_{\rho \sigma \,} (l+ k_2, k_2') \\
        & \times V^{\text{spin-0(1)}}_{\psi \theta \, } (k_2,l+k_2) \cdot \frac{-i}{(l+k_2)^2 } = - \frac{520 G^2 \xi  i}{3} q^4 \ln q^2.
    \end{split}
    \label{seagullampli00}
\end{eqnarray}
The corresponding bending angle becomes 
\begin{eqnarray}
        \theta^{\text{spin-0-spin-0}}_{\text{seagull}} =& \dfrac{8320 G^2 \xi }{E^2 \pi  b^7 M}\cdot
    \label{bendingseagull00}
\end{eqnarray}

\subsection*{\small  e) The double seagull diagrams :}
Let us now come to the double seagull diagrams given by the third of \ref{triangle2grav}.
The Feynman amplitude reads, 
\begin{eqnarray}
    \begin{split}
        i \mathcal{M}^{\text{spin-0-spin-0}}_{\text{double seagull}} =&  \frac{1}{2!} \int \frac{d^{\color{black}{d}} l}{(2 \pi)^{\color{black}{d}}} V_{\eta \lambda \rho \sigma(\xi)}^{\text{spin-0}(2)} (l,l+q) \frac{-i \mathcal{P}^{ \rho \sigma \mu \nu }}{(l+q)^2} \frac{-i \mathcal{P}^{\eta \lambda \alpha \beta}}{l^2} V^{\text{spin-0}(2)}_{ \alpha \beta \mu \nu \, } (k_2,k_2 ',m)  \\ 
         =& \frac{20  G^2 \xi i}{3}  q^4 \ln q^2.
    \end{split}
\end{eqnarray}
Its contribution to the angle of bending reads,
\begin{equation}
        \theta^{\text{spin-0-spin-0}}_{\text{double seagull}} = -\dfrac{320 G^2 \xi }{E^2 \pi  b^7 M}\cdot
    \label{potamp2}
\end{equation}
\subsection*{\small f) The fish diagrams :}
There are  two fish diagrams given by the  fourth and fifth of \ref{triangle2grav}. The corresponding Feynman amplitudes read,
\begin{eqnarray}
    \begin{split}
          i \mathcal{M}^{\text{spin-0-spin-0}}_{\text{fish-1}} =& \frac{1}{2!} \int \frac{d^{\color{black}{d}} l}{(2 \pi)^{\color{black}{d}}} V^{\text{spin-0(2)}}_{\rho \sigma \psi \theta(\xi)} (l,l+q) \frac{-i \mathcal{P}^{ \rho \sigma \gamma \delta}}{l^2} \frac{-i \mathcal{P}^{\psi \theta \alpha \beta }}{(l+q)^2}  V^{\mu \nu(3)} _{\alpha \beta \gamma \delta} (l+q,q)  \frac{-i \mathcal{P}_{\mu \nu \lambda \phi }}{q^2} V_{\text{spin-0(1)}}^{\lambda \phi \,} (k_2,k_2')  \\ 
            =& \frac{1030   G^2 \xi i}{3}  q^4 \ln q^2,
    \end{split}
\end{eqnarray}
and 
\begin{eqnarray}
    \begin{split}
         i \mathcal{M}^{\text{spin-0-spin-0}}_{\text{fish-2}} =& \frac{1}{2!} \int \frac{d^{\color{black}{d}} l}{(2 \pi)^{\color{black}{d}}} V_{\text{spin-0}(1)}^{\lambda \phi(\xi)} (q) \frac{-i \mathcal{P}_{\lambda \phi \mu \nu}}{q^2} V^{\mu \nu(3)} _{\alpha \beta \gamma \delta} (l,-q) \frac{-i \mathcal{P}^{\alpha \beta \psi \theta}}{l^2}  \frac{-i \mathcal{P}^{\gamma \delta \rho \sigma}}{(l+q)^2} V^{\text{spin-0}(2)}_{\psi \theta \rho \sigma} (k_2,k_2')  \\
        =& -32 i G^2 \xi  q^4 \ln q^2.
    \end{split}
\end{eqnarray}
The corresponding bending angles read
\begin{equation}
        \theta^{\text{spin-0-spin-0}}_{\text{fish-1}} = -\frac{16480 G^2 \xi }{E^2 \pi  b^7 M}, \qquad  \theta^{\text{spin-0-spin-0}}_{\text{fish-2}} =\frac{1536 G^2 \xi }{E^2 \pi  b^7 M}\cdot
    \label{potamp2'}
\end{equation}

\subsection*{\small g) The vacuum polarisation diagram :}
Finally, we come to the vacuum polarisation diagram  shown in the last of \ref{triangle2grav}. The Feynman amplitude is given by, 
\begin{eqnarray}
           i\mathcal{M}^{\text{spin-0-spin-0}}_{\text{vac.pol.}} = V^{\text{spin-0}(1)}_{\mu \nu\,(\xi)} (q)  \dfrac{ - i \mathcal{P}^{\mu \nu \rho \sigma }}{q^2} \Pi_{\rho \sigma \lambda \phi } (q) \ \dfrac{- i \mathcal{P}^{\lambda \phi \gamma \delta}}{q^2}   V^{\text{spin-0(1)}}_{\gamma \delta} (k_2,k_2',m)  
            = -12 G^2 \xi i q^4 \ln q^2,
\end{eqnarray} 
where $ \Pi_{\alpha \beta \gamma \delta }$ is the gauge invariant one loop graviton self energy due to itself  after adding the ghost contribution, reading~\cite{tHooft:1974toh},
\begin{eqnarray}
\begin{split}
            \Pi_{\alpha \beta \gamma \delta } =& -\frac{2G}{\pi}\ln q^2 \Big[\frac{21}{120}q^4 I_{\alpha \beta \gamma \delta} + \frac{23}{120}q^4 \eta_{\alpha \beta}\eta_{\gamma \delta}- \frac{23}{120}q^2 (\eta_{\alpha \beta}q_{\gamma} q_{\delta} + \eta_{\gamma \delta} q_{\alpha} q_{\beta}) - \frac{21}{240}q^2 (q_{\alpha} q_ {\delta}\eta_{\beta \gamma} + q_{\beta} q_ {\delta}\eta_{\alpha \gamma} \\
            & + q_{\alpha} q_ {\gamma}\eta_{\beta \delta} + q_{\beta} q_ {\gamma}\eta_{\alpha \delta}) + \frac{11}{30}q_{\alpha} q_{\beta} q_{\gamma}q_{\delta}\Big].
        \end{split}
        \label{vpol}
\end{eqnarray}
The contribution to the angle of bending reads,
\begin{eqnarray}
        \theta^{\text{spin-0-spin-0}}_{\text{vac.pol.}} = \frac{576 G^2 \xi }{E^2 \pi  b^7 M}\cdot
    \label{potamp2vp}
\end{eqnarray}
%
\subsection{\small The full result :}
Combining now Eqs.~\ref{potamp21}, \ref{bendingseagull00}, \ref{potamp2},  \ref{potamp2'}, \ref{potamp2vp}, the total bending angle for massive and massless spin-0 fields at the leading order ${\cal O}(G^2\xi)$ becomes
\begin{equation}
         \theta^{\text{spin-0-spin-0}}_{\text{total}}= -\frac{7912 G^2 \xi }{E^2 \pi  b^7 M}\cdot
    \label{potamp2''}  
\end{equation} 
%

\section{Massive spin-0-massless spin-1 scattering}\label{proca}
In this section we wish to consider the scattering  between a massive spin-0 field and the photon, in order find out the bending angle  of the photon at ${\cal O}(G^2 \xi)$.
\subsection*{\small a) The triangle diagram :}
 The Feynman amplitude for the first of \ref{triangle2grav} reads
\begin{eqnarray}
    \begin{split}
        i \mathcal{M}^{\text{spin-0-spin-1}}_{\text{Triangle}} =& \int \frac{d^{\color{black}{d}} l}{(2 \pi)^{\color{black}{d}}} V^{\text{spin-0}(2)}_{\rho \sigma \psi \theta \,(\xi)} (l,l-q) \frac{{- i \cal P}^{\psi \theta \mu \nu }}{l^2} \frac{{- i \cal P}^{ \rho \sigma \alpha \beta}}{(l-q)^2} V^{\text{spin-1}(1)}_{\tau, \chi, \mu \nu} (k_2,l + k_2) \epsilon^{\tau}(k_2)\\
        & \times V^ {\text{spin-1}(1)}_{\phi, \zeta, \alpha \beta} (l+ k_2,k_2') \epsilon^{\star \zeta}(k'_2) \frac{-i \eta^{\chi \phi}}{(l+k_2)^2} 
        = \frac{2G^2 \xi i}{3}  q^4 \ln q^2 \left[49 \vec{\epsilon} \cdot \vec{\epsilon'}^{\star}  + 13  \hat{q}\cdot\vec{\epsilon} \ \hat{q}\cdot\vec{\epsilon'}^{\star} \right],
    \end{split}
    \label{trv}
\end{eqnarray}
where $\e(k_2)$, $\e^{\star}(k'_2)$ respectively stand for polarisation vectors representing annihilation and creation of photons.
The corresponding bending angle is given by,
\begin{eqnarray}
        \theta^{\text{spin-0-spin-1}} _{\text{triangle}}(G^2\xi) = -\frac{24 G^2 \xi}{E^2 b^7 M \pi} \left[ 61\vec{\epsilon} \cdot \vec{\epsilon}'^{\star} +26  \vec{\epsilon}'^{\star}\cdot \hat{b}\  \vec{\epsilon}\cdot\hat{b}\right].
        \label{tr}
\end{eqnarray}
\subsection*{\small  b) The seagull diagram :}
 The Feynman amplitude for the second of \ref{triangle2grav} reads,
\begin{eqnarray}
    \begin{split}
        i \mathcal{M}^{\text{spin-0-spin-0}}_{\text{seagull}} =& \int \frac{d^{\color{black}{d}} l}{(2 \pi)^{\color{black}{d}}} V_{\text{spin-0}(1)}^{\lambda \phi\,(\xi)} (q) \frac{-i \mathcal{P}_{\lambda \phi \mu \nu}}{q^2} V^{\mu \nu (3)} _{\alpha \beta \gamma \delta} (l-q,-q) \frac{-i \mathcal{P}^{\alpha \beta \psi \theta}}{l^2}  \frac{-i \mathcal{P}^{\gamma \delta \rho \sigma}}{(l-q)^2}   V^{\text{spin-1}(1)}_{\tau, \chi, \psi \theta \, } (k_2,l + k_2) \epsilon^{\tau}(k_2)\\
        & \times V^{\text{spin-1}(1)}_{\phi, \zeta, \rho \sigma} (l+ k_2,k_2') \epsilon^{\star \zeta}(k'_2) \frac{-i \eta^{\chi \phi}}{(l+k_2)^2} 
        =   \dfrac{8 G^2 q^2 \xi i}{3}  
        q \cdot \epsilon \ 
        q \cdot {\epsilon'}^{\star}
        \ln q^2
        -174\ G^2 q^4 \xi i \
        \epsilon \cdot {\epsilon'}^{\star}
        \ln q^2,
    \end{split}
    \label{seagullampli01}
\end{eqnarray}
and the corresponding bending angle becomes, 
\begin{eqnarray}
        \theta^{\text{spin-0-spin-0}}_{\text{seagull}} = - \dfrac{64 G^2\xi}{b^7 E^2 M \pi} \left[3 \hat{b} \cdot \vec{\epsilon} \ \hat{b} \cdot\vec{\epsilon'}^{\star}
- 128\vec{\epsilon}\cdot \vec{\epsilon'}^{\star}\right].
    \label{bendingseagull01}
\end{eqnarray}
\subsection*{\small c) The double seagull diagram :}
The Feynman amplitude for the third diagram of \ref{triangle2grav} is given by,
\begin{eqnarray}
    \begin{split}
         i \mathcal{M}^{\text{spin-0-spin-1}}_{\text{double seagull}} =& \frac{1}{2!} \int \frac{d^{\color{black}{d}} l}{(2 \pi)^{\color{black}{d}}} V_{\eta \lambda \rho \sigma \,(\xi)}^{\text{spin-0}(2)} (l+q,l) \frac{-i \mathcal{P}^{ \rho \sigma \mu \nu }}{(l+q)^2} \frac{-i \mathcal{P}^{\eta \lambda \rho \sigma}}{l^2}   V^{\text{spin-1(2)}}_{\beta,\alpha ,\mu \nu \rho \sigma} (k_2,k_2 ' ) \epsilon^{\beta} \epsilon^{\star  \alpha} \\
         &=-274 G^2 \xi  i q^4 \ln q^2 \vec{\epsilon} \cdot \vec{\epsilon}^{'\star} -\frac{20 G^2 \xi i}{3}  \, q^2 \ln q^2 \vec{q}\cdot \vec{\epsilon} \ \vec{q}\cdot \vec{\epsilon'}^{\star}.
    \end{split}
\end{eqnarray}
The bending angle is given by,
\begin{eqnarray}
    \theta_{\text{double seagull}}^{\text{spin-0-spin-1}} = \frac{16 G^2 \xi}{E^2   b^7 M \pi}   \left( 817 \vec{\epsilon} \cdot \vec{\epsilon'}^{\star}+30  \vec{\epsilon'}^{\star}\cdot \hat{b} \ \vec{\epsilon}\cdot \hat{b}  \right).
        \label{v3}
\end{eqnarray}

\subsection*{\small d) The fish diagrams :}
There are two fish diagrams for this scattering process given by the fourth and fifth of \ref{triangle2grav}. The Feynman amplitudes for them respectively read, 
\begin{eqnarray}
    \begin{split}
        i \mathcal{M}^{\text{spin-0-spin-1}}_{\text{fish-1}} =& \frac{1}{2!} \int \frac{d^{\color{black}{d}} l}{(2 \pi)^{\color{black}{d}}} V^{\text{spin-0}\,(2)}_{\rho \sigma \psi \theta \,(\xi)} (l,l+q) \frac{-i \mathcal{P}^{ \rho \sigma \gamma \delta }}{l^2} \frac{-i \mathcal{P}^{ \psi \theta \alpha \beta}}{(l+q)^2} V^{\mu \nu \,(3)} _{\alpha \beta \gamma \delta} (l+q,q) \frac{-i \mathcal{P}_{ \mu \nu\lambda \phi }}{q^2} \\
        & \times V^{\tau, \chi, \lambda \phi}_{\text{spin-1\,(1)}} (k_2,k_2') \epsilon_{\tau}(k_2) \epsilon^{\star}_{\chi}(k'_2) 
        = \frac{620}{3} G^2 \xi \, q^2 \ln q^2 \vec{q}\cdot \vec{\epsilon}  \ \vec{q}\cdot \vec{\epsilon'}^{\star},
    \end{split}
\end{eqnarray}
and,
\begin{eqnarray}
    \begin{split}
        i \mathcal{M}^{\text{spin-0-spin-1}}_{\text{fish-2}} =& \frac{1}{2!} \int \frac{d^{\color{black}{d}} l}{(2 \pi)^{\color{black}{d}}} V_{\text{spin-0}\,(1)}^{\lambda \phi\,(\xi)} (q) \frac{-i \mathcal{P}_{ \lambda \phi \mu \nu }}{q^2}  V^{\mu \nu \,(3)} _{\alpha \beta \gamma \delta} (l,-q) \frac{-i \mathcal{P}^{ \gamma \delta \rho \sigma  }}{(l+q)^2} \frac{-i \mathcal{P}^{ \alpha \beta \psi \theta }}{l^2}  \\
        & \times V_{\tau, \chi,\rho \sigma \psi \theta }^{\text{spin-1\,(2)}} (k_2,k_2') \epsilon^{\tau} (k_2)\epsilon^{\star \chi} (k'_2) = 32 G^2 \xi i \ q^2 \ln q^2 \vec{q}\cdot \vec{\epsilon} \ \vec{q}\cdot \vec{\epsilon'}^{\star}+1560 G^2 \xi i q^4 \ln q^2 \vec{\epsilon} \cdot\vec{\epsilon'}^{\star}.
    \end{split}
\end{eqnarray}
The bending angles respectively, are given by,
\begin{eqnarray}
   &&     \theta^{\text{spin-0-spin-1}}_{\text{fish-1}} = \frac{2480 G^2 \xi }{E^2  b^7 M \pi}\left(\vec{\epsilon} \cdot \vec{\epsilon'}^{\star} -6\vec{\epsilon'}^{\star}\cdot\hat{b} \ \vec{\epsilon}\cdot \hat{b}\right), \quad \theta^{\text{spin-0-spin-1}}_{\text{fish-2}} = -\frac{768 G^2 \xi }{E^2  b^7 M \pi}\left( 97\vec{\epsilon} \cdot \vec{\epsilon'}^{\star} +3\vec{\epsilon'}^{\star}\cdot\hat{b} \ \vec{\epsilon}\cdot \hat{b}\right). 
        \label{v45}
\end{eqnarray}
\subsection*{\small e) The vacuum polarisation diagram :}
The Feynman amplitude for the last  of \ref{triangle2grav} is given by,
\begin{eqnarray}
    \begin{split}
        \mathcal{M}^{\text{spin-0-spin-1}}_{\text{vac-pol}} =& V^ {\text{spin-0}\,(1)}_{\mu  \nu \,(\xi)} (q) \dfrac{ - i \mathcal{P}^{\mu \nu \rho \sigma }}{q^2} \Pi_{\rho \sigma \lambda \phi } (q) \ \dfrac{- i \mathcal{P}^{\lambda \phi \gamma \delta}}{q^2}   V_{\beta ,\alpha, \gamma \delta}^{\text{spin-1\,(1)}} (k_2,k_2') \epsilon^{\beta}(k_2) \epsilon^{\star  \alpha}(k'_2) 
        = -24 G^2 \xi i \ q^2 \ln q^2 \hat{q}\cdot \vec{\epsilon}  \ \hat{q}\cdot \vec{\epsilon'}^{\star},
    \end{split}
\end{eqnarray}  
 where  $\Pi_{\rho \sigma \lambda \phi } (q)$ is as earlier the gauge invariant one loop graviton self energy due to itself, 
 Eq.~\ref{vpol}~\cite{tHooft:1974toh}. The corresponding bending angle is given by,
\begin{eqnarray}
    \theta^{\text{spin-0-spin-1}}_{\text{vac-pol}} = - \frac{288 G^2 \xi}{E^2  b^7 M \pi}\left(\vec{\epsilon} \cdot \vec{\epsilon'}^{\star} -6 \vec{\epsilon'}^{\star}\cdot\hat{b} \ \vec{\epsilon}\cdot\hat{b}\right)\cdot
     \label{v6}
\end{eqnarray}
\subsection{\small The full result :}
Combining now the individual contributions from Eqs.~\ref{tr}, \ref{bendingseagull01}, \ref{v3}, \ref{v45} and \ref{v6}, we obtain the  bending angle between a massive spin-0 and massless spin-1 field at the leading ${\cal O}(G^2 \xi)$,
\begin{eqnarray}
        \theta^{\text{spin-0-spin-1}}_{G^2\xi}=- \frac{8G^2 \xi}{E^2   b^7 M \pi} \left(7611 \vec{\epsilon} \cdot \vec{\epsilon'}^{\star} +902\vec{\epsilon'}^{\star}\cdot\hat{b} \ \vec{\epsilon}\cdot\hat{b} \right)\cdot
    \label{total}
\end{eqnarray}

\section{Discussion}\label{disc}
We have derived in this paper the leading effect of $\sqrt{-g}\xi R \phi^2/2$ non-minimal coupling for a massive scalar $\phi$, on the scattering of massless scalar and photons. The  angle of light bending at ${\cal O}(G^2 \xi)$ is given by Eqs.~\ref{potamp2''}, \ref{total}, which are the main results of this paper. Given that such coupling naturally arises for the sake of  renormalisability when gravity is turned on~\cite{Parker:2009uva}, we believe this result is interesting in its own right, even though this bending must be tiny compared to the minimal ones. \\

\noindent
We note that for $\xi {\ensuremath >}0$, the non-minimal bending angle is negative, indicating an effective repulsive effect.
Next, taking $b\sim L/E$ for the impact parameter, it is clear that the same is proportional to $E^5$, in qualitative agreement with the minimal results, which behave as powers of $GM/b$. \\

\noindent
{\color{black}{Let us now see how our results compare with the lowest order minimal expression, $\theta_0 = 4GM/b$. We consider the example of the Sun and visible light. We take $M\simeq1.99\times10^{30}\,\mathrm{kg}$, $b=R_\odot\simeq6.957\times10^8\,\mathrm{m}$ for  rays grazing through the surface of the Sun.  We have the wavelengths, $\lambda_M=h/(Mc)\simeq1.769\times10^{-73}\,\mathrm{m}$ and $\lambda_\gamma\simeq 8.0\times10^{-8}\,\mathrm{m}$.
		The leading deflection is $\theta_{0}\simeq8.49\times10^{-6}\,\mathrm{rad}$, whereas for our case  we have $|\theta_{\mathrm{NMC}}|\sim9.69\times10^{-289}|\xi|\,\mathrm{rad}$.
		Thus $|\theta_{\mathrm{NMC}}|/\theta_{0}\sim {\cal O}(10^{-283}|\xi|)$, which is  much too small compared unity for any $|\xi|\lesssim {\cal O}(1)$.}}\\

Finally, we emphasise the peculiar  dependence of the non-minimal bending angles on the  mass $M$ of the gravitating object, Eqs.~\ref{potamp2''}, \ref{total},  indicating an increase of the angle with the decrease of the mass, while the other things are held fixed. This feature seems to be very counter-intuitive, and it originates from the fact that the non-minimal vertices do not contain any explicit momenta of the scalar field, Eq.~\ref{qg21a}. Does this feature  rule out $\xi$ completely? Or does it  indicate  the possibility of dependence of $\xi$ on $M$ in a dimensionless manner? In other words, is it somehow possible that $\xi$ increases with the increase of the mass of the gravitating object? Perhaps an RG flow analysis for $\xi$ can give us further insight into this, and  we hope to come back to this issue in future work.

 
\bigskip
\bigskip
\noindent
{\bf Acknowledgements :} AKN's research is supported by the research fellowship of the University Grants Commission, Govt. of India (NTA Ref. No./Student ID: 221610099618). ASM’s research is supported by the
fellowship of the State Government Departmental Fellowship Scheme (letter no: P1/RS/58/26), Govt. of West Bengal (India).  The authors thank  anonymous referee for  some constructive critical comments on an earlier version of this manuscript.


\end{document}